\documentclass[12pt]{article}

\usepackage{graphicx, fullpage, amsmath, amssymb}
\usepackage{empheq}
\usepackage{color}
\usepackage[colorlinks=true,linkcolor=blue,citecolor=magenta]{hyperref}
\usepackage{lscape}
\usepackage{nicefrac}
\usepackage[T1]{fontenc}
\usepackage[utf8]{inputenc}
\usepackage{authblk}
\usepackage{xr}
\usepackage{comment}
\usepackage{tikz}
\usepackage{float}
\usepackage[font={footnotesize},margin=1in]{caption}
\usepackage{scicite}
\usepackage{placeins}

\newcommand*\circled[1]{%
	\protect\tikz[baseline=(char.base)]{
		\scriptsize
		\protect\node[shape=circle,draw,inner sep=1.5pt] (char) {#1};
	}
}
\newcommand\blfootnote[1]{%
  \begingroup
  \renewcommand\thefootnote{}\footnote{#1}%
  \addtocounter{footnote}{-1}%
  \endgroup
}

\begin{document}

\title{\LARGE \bf
Reducing Cascading Failure Risk by Increasing Infrastructure Network Interdependency
}
\author[1]{\rm Mert~Korkali}
\author[2]{\rm Jason~G.~Veneman}
\author[2,3]{\rm Brian~F.~Tivnan}
\author[3,4,*]{\rm Paul~D.H.~Hines}
\affil[1]{\small Lawrence Livermore National Laboratory, Livermore,
CA 94550 USA}
\affil[2]{The MITRE Corporation, McLean, VA 22102 USA}
\affil[3]{Vermont Complex Systems Center, The University of Vermont, Burlington,
VT 05405 USA}
\affil[4]{School of Engineering, The University of Vermont, Burlington,
VT 05405 USA}
\vspace{.1in}
\affil[*]{To whom correspondence should be addressed; E-mail: paul.hines@uvm.edu}

\date{}

\maketitle

\blfootnote{\copyright The University of Vermont and the MITRE Corporation. All rights reserved.\\ Approved for Public Release; Distribution Unlimited. 14-3504}

\begin{abstract}
Increased coupling between critical infrastructure networks, such as power and communication systems, will have important implications for the reliability and security of these systems. 
To understand the effects of power-communication coupling, several have studied interdependent network models and reported that increased coupling can increase system vulnerability. However, these results come from models that have substantially different mechanisms of cascading, relative to those found in actual power and communication networks. This paper reports on two sets of experiments that compare the network vulnerability implications resulting from simple topological models and models that more accurately capture the dynamics of cascading in power systems. First, we compare a simple model of topological contagion to a model of cascading in power systems and find that the power grid shows a much higher level of vulnerability, relative to the contagion model. Second, we compare a model of topological cascades in coupled networks to three different physics-based models of power grids coupled to communication networks. Again, the more accurate models suggest very different conclusions. In all but the most extreme case, the physics-based power grid models indicate that increased power-communication coupling decreases vulnerability. This is opposite from what one would conclude from the coupled topological model, in which zero coupling is optimal. Finally, an extreme case in which communication failures immediately cause grid failures, suggests that if systems are poorly designed, increased coupling can be harmful. Together these results suggest design strategies for reducing the risk of cascades in interdependent infrastructure systems.
\end{abstract}

\normalsize

\section*{Introduction}

Understanding the reliability and security implications of increased coupling between interdependent power, water, transportation and communication infrastructure systems is critical, given the vital services that these infrastructures provide and continuing threats posed by natural disasters and terrorist attacks~\cite{Brummittetal13,Rinaldietal01}.
This is particularly true for the coupling between electric power and communications networks, given 
the essential nature of electric power to modern societies, 
the rapid growth of smart grid technology~\cite{Morgan:2009}, and 
the potential for cascading failure to lead to catastrophic blackouts~\cite{Dobsonetal07}.
Smart grid systems, such as Advanced Metering Infrastructure and microprocessor-based controls, can be valuable tools for mitigating these risks~\cite{EPRIReport09}.
But automation can also introduce new failures mechanisms: 
cyber-attacks may reach a larger number of critical components~\cite{Wei:2011} and 
outages may propagate between the connected networks, increasing the risk of massive failures.


In order to quantify the risks and benefits of network interdependency, models are needed that at least approximately represent the potential for cascading within a power grid, as well as between power and communication networks.
A variety of models have been suggested for understanding the mechanisms by which failures, ideas, and diseases 
propagate within individual networks~\cite{Bailey75,Watts02,Leskovecetal09}.
Simple models clearly show that different types of networks can respond very differently to random failures and volitional attacks~\cite{Albertetal00,Magnienetal11,Asztalosetal14}. 
Subsequently, several have suggested that contagion-style models be used to understand vulnerability in power grids~\cite{MotterLai02,ChassinPosse05,XiaoYeh11,Schneideretal13}.

\begin{figure}[ht]
\centering \includegraphics[width=.5\columnwidth]{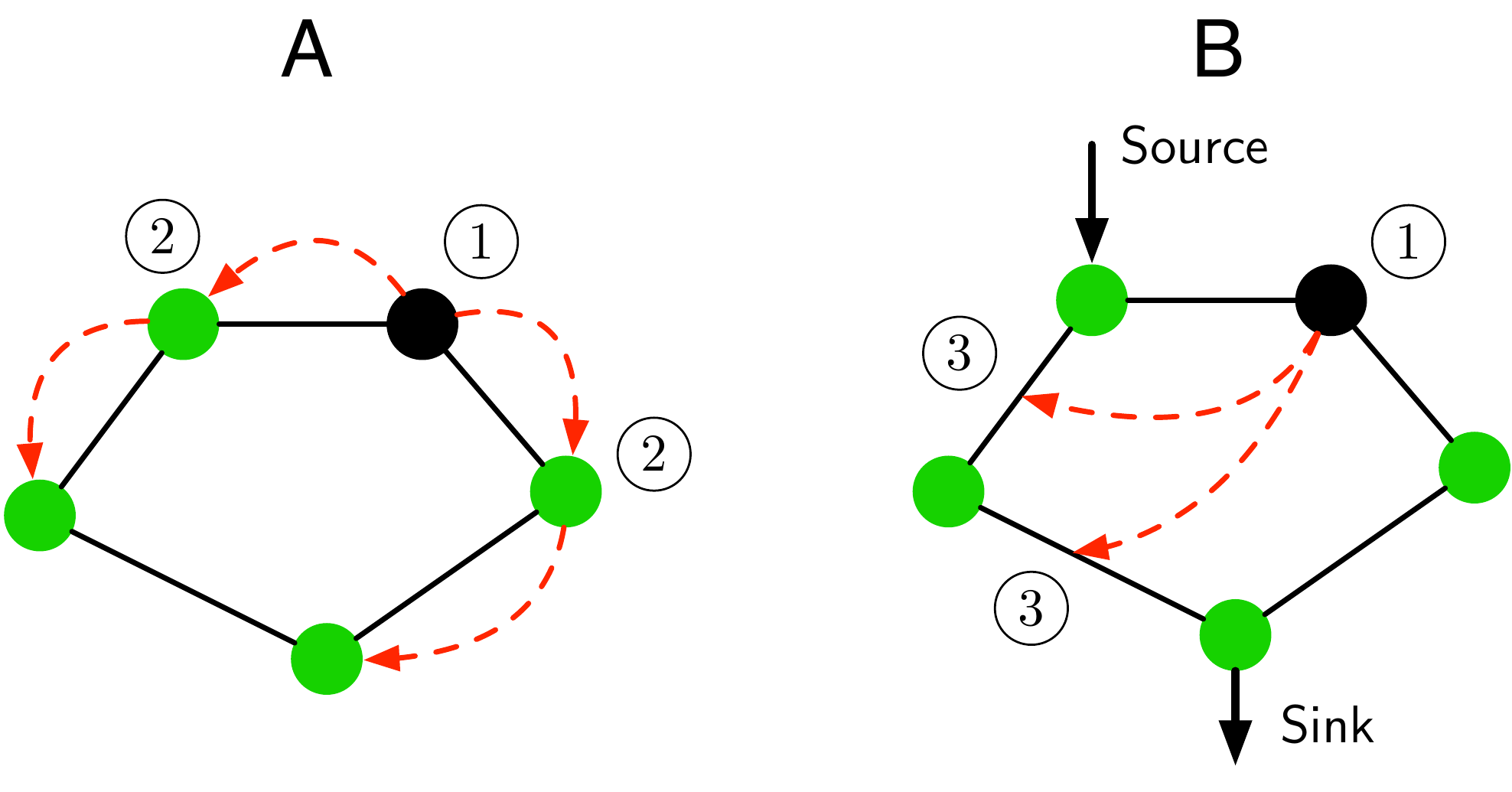}
\caption{%
Comparative illustration of cascade propagation in 
(A) topological contagion and 
(B) power grid models. 
In topological models of cascading, such as the contagion model in \cite{Watts02} or the sandpile model from \cite{Brummittetal12}), cascades propagate from the initiating failure \circled{1}~to neighboring nodes \circled{2}. 
In a power grid, the initiating failure \circled{1}~causes 
increased loads along parallel paths~\circled{3}, 
which may subsequently fail~\cite{Hinesetal10}.
}
\label{fig:Topology_only}
\end{figure}

However, power grids differ in important ways from these simple models.
In a contagion-style model, failures propagate locally:
when component~$i$ fails, the next component to fail is topologically connected to~$i$.
On the other hand, power grids are 
engineered networks, in which energy flows from generators to loads through power lines (edges), each of which has a limit on the amount of electrical flow it can tolerate. 
When node (substation) or edge (transmission line) failures occur, power re-routes according to Kirchhoff's and Ohm's laws.
This re-routing increases flows along parallel paths, which can subsequently trigger long chains of component failures,
potentially leading to a wide-area blackout~\cite{Dobsonetal07}. 
As a result of this process, failures propagate non-locally: the next component to fail may be hundreds of miles or tens of edges distant from the previous failure.
Thus, overly simple topological models can lead to misleading conclusions~\cite{Hinesetal10} (Figure~\ref{fig:Topology_only}).


On the other hand, simple models can often suggest new approaches to a particular problem, particularly when there is limited existing understanding, as is the case with vulnerability in interdependent networks.
Motivated, at least in part, by increasing interdependency between power and communications networks, a number of recent studies suggest that interdependency can increase vulnerability in network structures that were otherwise relatively robust~\cite{Buldyrevetal10,Bashanetal13,Gaoetal11,Nguyenetal13}.
Others have found non-monotonic relationships between the level of coupling between interdependent networks and network performance, suggesting that there exists an optimal level of coupling~\cite{Schneideretal13,Brummittetal12}. 
More recent results suggest that under some conditions, coupling between networks can improve performance~\cite{reis2014avoiding}.
While these results clearly show that coupling is important to the performance of interdependent systems, the ``typical'' impact of coupling is not clear.
More work is needed to understand the conditions under which coupling is beneficial, or harmful.

Again, the results above come from models that diverge from real infrastructure networks in important ways, making it hard to understand the implications for a particular system of interest.
First, the topological structures found in infrastructure networks differ notably from standard abstract models~\cite{DuenasOsorioVemuru09,CotillaSanchezetal12}, 
largely due to geographic and cost constraints~\cite{Gastner:2006}.
Second, the physical mechanisms of cascading within networks (see Figure~\ref{fig:Topology_only}) and between interdependent networks (see Figure~\ref{fig:coupled_models}) are notably different from those of the percolation-style models in~\cite{Buldyrevetal10,Brummittetal12,Chenetal13}. 
Recent results suggest that modeling the physics of power flows can have important impacts on the conclusions that one would draw from interdependent infrastructure network models~\cite{RahnamayHayat13,Parandehgheibietal14}.
In order to understand the extent to which insights from abstracted network models can be useful for particular types of interdependent networks (such as power and communications networks), comparisons are needed between simple models and those that more accurately capture the topology, physics, and coupling of particular infrastructure systems.

Therefore, the goal of this paper is to understand the impact of 
network topology, cascading mechanisms (physics) and coupling on infrastructure network vulnerability.
We use the case of increased coupling between electric power systems and communication networks (Smart Grid) as an illustrative test case.
Two sets of numerical experiments combine to address this goal. 
The first set of experiments focuses on topology and physics.
In this experiment we compare the relative vulnerability of different topological structures to random disturbances 
given two different models of intra-network cascading: 
a simple contagion model and a model that more accurately captures the mechanisms of cascading in power grids.
The second set of experiments compares the impact of increased inter-network interdependency on vulnerability, given different models cascading failures propagation.



\section*{Results}

The two sets of results described here describe the vulnerability of different network structures with different models of cascading to random node failures of various sizes.
Each of the networks was sized to have~$n=2383$ nodes and~$m=2886$ links to correspond to the size of our power grid test system (a model of the Polish system~\cite{Zimmermanetal11}).
In each of our experiments we vary the size of the initiating failure~$f$, which is the ratio of the number of nodes in the initial random failure to the total number of nodes in the network,~$n$.
The ultimate impact of each cascade is measured by the number of nodes remaining within the largest (giant) connected component of the graph,~$|\text{GC}|$, after the cascade has subsided.
We estimated the vulnerability of each network to initiating failures of different sizes by measuring the probability that the largest connected connected component in the post-cascade network,~$GC$, includes more than half of the nodes, i.e.,~$\Pr(|\text{GC}|>0.5n)$.


\subsection*{Intra-network cascading}

Our first set of experiments compares the vulnerability of five different network structures 
(a power grid, a square lattice, an Erd\H{o}s-R\'{e}nyi random graph, a random regular network, and a scale-free network)
using the two different models of cascade propagation illustrated in Figure~\ref{fig:Topology_only}.

\begin{figure}[t]
\begin{center}
\centerline{\includegraphics[width=3.5in]{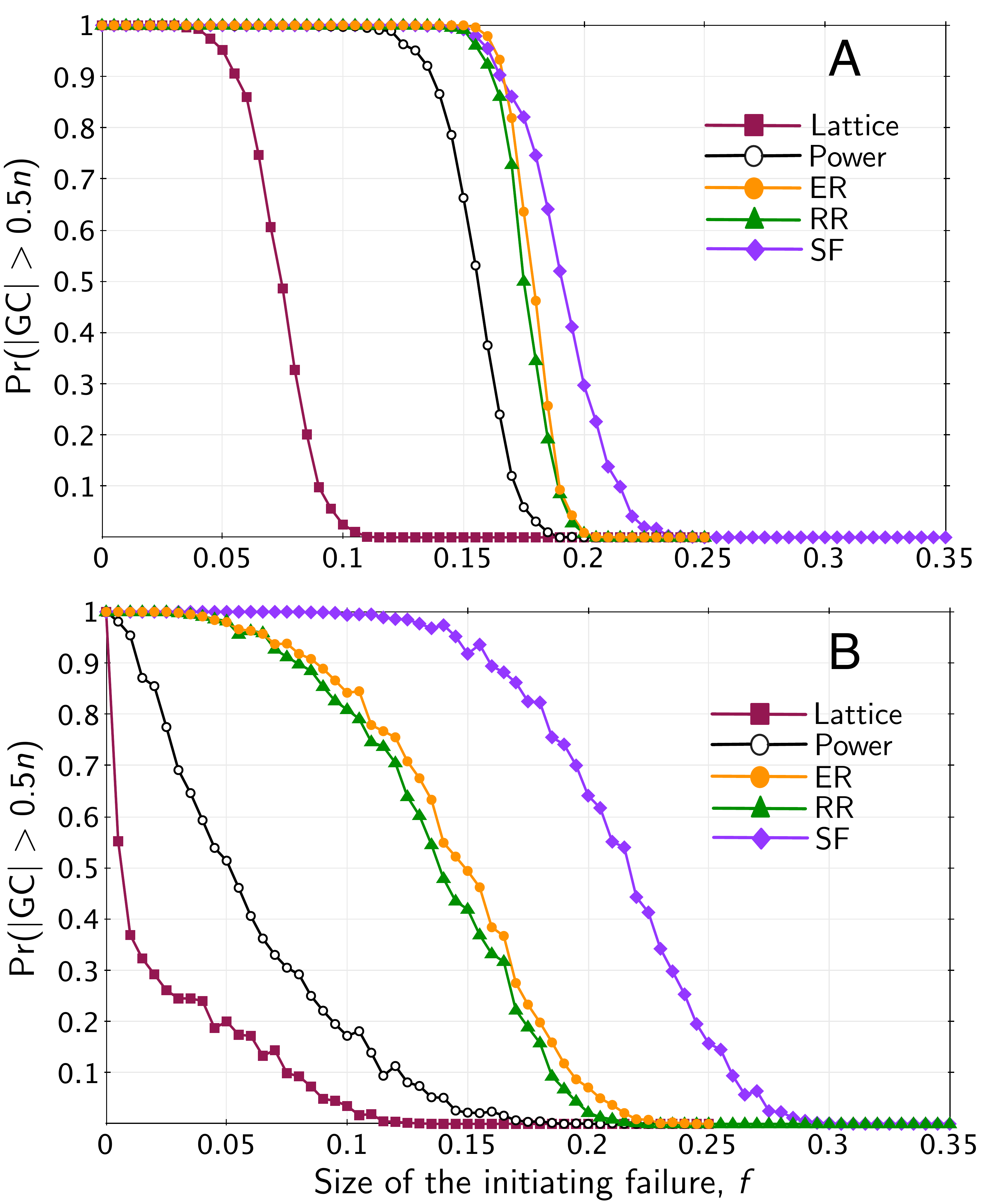}}
\caption{Robustness of several networks structures to random node failures, measured by the probability that the post-cascade giant component includes at least 50\% of the network, in (A) a Topological Contagion model and in a (B) Power Grid cascading model.}
\label{fig:GC_figures_SingleNet}
\end{center}
\end{figure}

Figure~\ref{fig:GC_figures_SingleNet}A shows results from our first model of cascading: 
a simple model of topological contagion, proposed by Watts in~\cite{Watts02}.
In this model, after the initial set of~$\sim fn$ node failures, 
Node~$i$ fails if the fraction of Node~$i$'s neighbors that are in a failed state exceeds some threshold~$\phi_i$.
In these results, each~$\phi_i$ was randomly drawn from a uniform distribution over~$(0,1)$.

Figure~\ref{fig:GC_figures_SingleNet}B shows the results from our second model, which more accurately represents the dynamics of cascading overloads in a power transmission network.
In this power grid model~\cite{EppsteinHines12}, 
the failure of edges results in the redistribution of power flows along parallel paths according to a linearized power flow model~(see Materials and Methods).
This new distribution of flows can cause edges to be overloaded,
possibly inducing further edge failures. 
If edge failures cause the network to fracture into separate connected components,
power sources (generators) and power sinks (loads) adjust to arrive at a new balance between supply and demand. 
Once started, cascades continue until no overloaded edges remain.


These results from these two models show some notable similarities. 
From both models of cascading, the power grid and lattice structures appears to be most vulnerable
and the scale-free topology is the most robust.
In fact, the relative order of the five networks is nearly identical in Figures~\ref{fig:GC_figures_SingleNet}A and \ref{fig:GC_figures_SingleNet}B.

On the other hand, the power grid model accentuates the vulnerability differences among the different topologies
and changes the nature of the transition in~$p$.
In the Power Grid model, we do not observe the rapid, second-order phase transition that is apparent in the topological model; 
transitions as~$f$ increases are more gradual.
Whereas the midpoint of the transition
is similar in the two models (power and topological) for the scale-free network, 
the Polish power grid and lattice structures appear to be much more vulnerable from the perspective of the Power Grid model.


\subsection*{Inter-network cascading}

\begin{figure}[H]
\centering \includegraphics[width=3.5in]{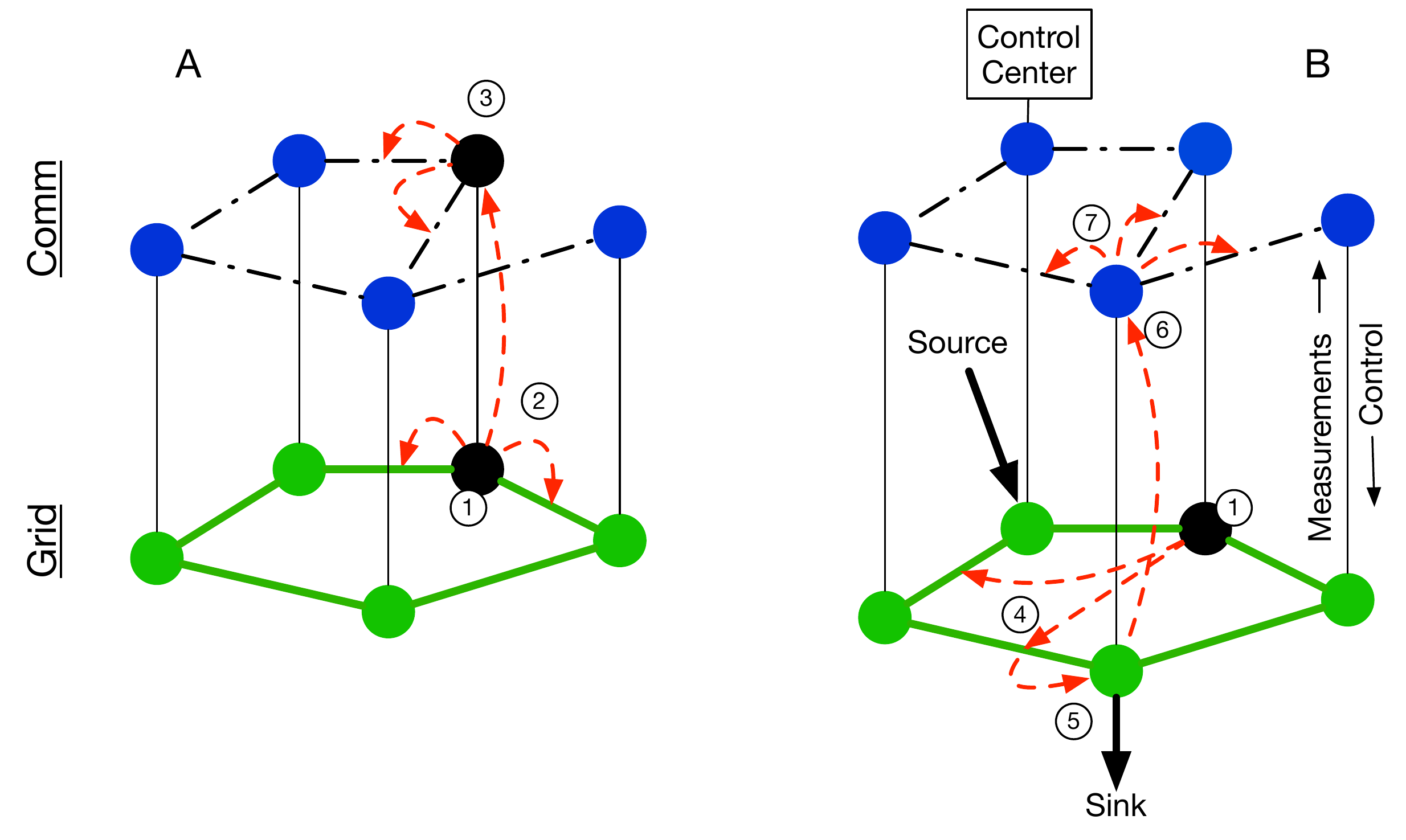}
\caption{%
Comparative illustration of the (A) ``Coupled Topological'' model and the (B) ``Non-ideal Smart Grid'' model. 
In the Coupled Topological model an initiating disturbance~\circled{1} causes \circled{2} edge failures in the power grid as well as \circled{3} node and edge failures in the communications (comm) network. 
As a result, the size of the giant component is reduced to~$0.8n$. 
In the Non-ideal Smart Grid (SG) model the initiating failure~\circled{1} potentially causes overloads \circled{4}, which causes an edge failure and \circled{5} a loss of power at the ``sink'' node.
This may (depending on the availability of backup power) cause a communication node failure \circled{6} and thus  
communication link failures \circled{7}, 
which fracture the communication network and prevent messages from being passed from and to the control center.
}
\label{fig:coupled_models}
\end{figure}

Our second set of experiments explores the impact of interdependency on network vulnerability.
Specifically, we considered a pair of interdependent networks (a power grid and a communication network, denoted hereafter by~$\mathcal{N}_P$ and~$\mathcal{N}_C$, respectively), in which a fraction~$q$ (degree of coupling) of the~$n$ nodes in~$\mathcal{N}_P$ are coupled to corresponding nodes in~$\mathcal{N}_C$. 
As in the first set of experiments, two different types of models are compared: one that is purely topological (Figure~\ref{fig:coupled_models}A) and a second that includes additional details about power flows (Figure~\ref{fig:coupled_models}B).

The first model is an implementation of the interdependent cascade/percolation model proposed in~\cite{Buldyrevetal10}.
When a node fails in this model, the associated edges in network~$\mathcal{N}_P$ and~$\mathcal{N}_C$ immediately fail. 
If the removed edges result in unconnected clusters in~$\mathcal{N}_P$ (or~$\mathcal{N}_C$),
then the edges linking the clusters in~$\mathcal{N}_C$ (or~$\mathcal{N}_P$) fail.
This cascading process continues until both~$\mathcal{N}_C$ and~$\mathcal{N}_P$ have the same set of clusters.
Henceforth, this model will be referred to as the ``Coupled Topological Model'' (see Figure~\ref{fig:coupled_models}A).

While it is clear that smart grid will result in some inter-network dependencies, 
it is not clear exactly what mechanisms of inter-network cascading will exist as interdependency increases.
Therefore, in a second set of coupled network models, we modeled three different possibilities for the nature of this coupling.
In all three ``Smart Grid'' models, cascades are allowed to propagate within the power grid, as in the previous model, 
with the exception that the communication network is used to collect measurements and issue control commands to the power grid.
In the smart grid models, if there is a~$\mathcal{N}_P \leftrightarrow \mathcal{N}_C$ connection at Node~$i$ and there is a valid path from~$i$ to the network's centrally located control center, 
then the control system is able to collect measurements from the network, such as data about the flows on overloaded transmission lines.
Similarly, sources or sinks at Node~$j$ can be controlled only if there is a valid path from~$j$ through~$\mathcal{N}_C$ to the control center. 
Now, instead of a component failing quickly after an overload occurs, measurements can be collected and used to choose optimal control actions (rapid reductions in nodal supply or demand) that could mitigate propagation of the cascade.
Once chosen these decisions are distributed through~$\mathcal{N}_C$ to the appropriate nodes in~$\mathcal{N}_P$.

In the first of three variants on this model, the ``Ideal Smart Grid,'' we assume that communication nodes continue to operate, even if nodes in~$\mathcal{N}_P$ fail. 
This corresponds to the case where~$\mathcal{N}_C$ has highly reliable battery backup systems that allow it to continue to operate when power failures occur, as is common practice in the design of modern SCADA (Supervisory Control and Data Acquisition) systems.

In our second variant, ``Non-ideal Smart Grid,'' communication nodes fail with a probability that is proportional to the amount of local load shedding.%
For example, if Node~$i$ in~$\mathcal{N}_P$ has had to shed 50\% of its local load,
Node~$i$ in~$\mathcal{N}_C$ will fail with a 0.5 probability. 
Since the possibility exists for communication node failures, nodes in~$\mathcal{N}_P$ will lose the ability to be monitored and controlled if there ceases to be a functional communications network path between the control center and a particular grid node.
If communication node/edge failures cause~$\mathcal{N}_C$ to fracture into clusters, signals can only pass within the cluster where the control center is located (see Figure~\ref{fig:coupled_models}).

Finally, in our third variant, the ``Vulnerable Smart Grid,'' generators and loads at node~$i$ fail immediately when there the corresponding communication Node~$i$ fails, if there is a~$\mathcal{N}_P \leftrightarrow \mathcal{N}_C$ connection at~$i$.
This is the most pessimistic of the three models, and diverges substantially from industry design standards which seek to minimize the chance that power failures will cause communication failures, and vice versa.

To build semi-realistic coupled network topologies, we used the data for the Polish power grid for~$\mathcal{N}_P$ and connected a fraction~$q$ of the~$n$ nodes to a communication network~$\mathcal{N}_C$.
Because both power and communication networks are geographically embedded,
$\mathcal{N}_P$ and~$\mathcal{N}_C$ are likely to be somewhat, but not perfectly, correlated.
To approximate this correlation,~$\mathcal{N}_C$ was initialized to be identical to~$\mathcal{N}_P$, and then 10\% of the edges in~$\mathcal{N}_C$ were randomly rewired.

After initializing the data and models, the various models were, as before, subjected to random node failures, and the performance of the networks measured.
For the Coupled Topological results, we measured network performance using the giant component probability~$\Pr(|\text{GC}|>0.5n)$.
For the power grid models, we measured both~$\Pr(|\text{GC}|>0.5n)$ and an analogous measure of performance: the probability that the network can serve at least 50\% of the load in the network, after the cascade has subsided,~$\Pr(P_T>0.5P_0)$.
(see Statistical Analysis).

\begin{figure}[ht]
\centering \includegraphics[width=.5\columnwidth]{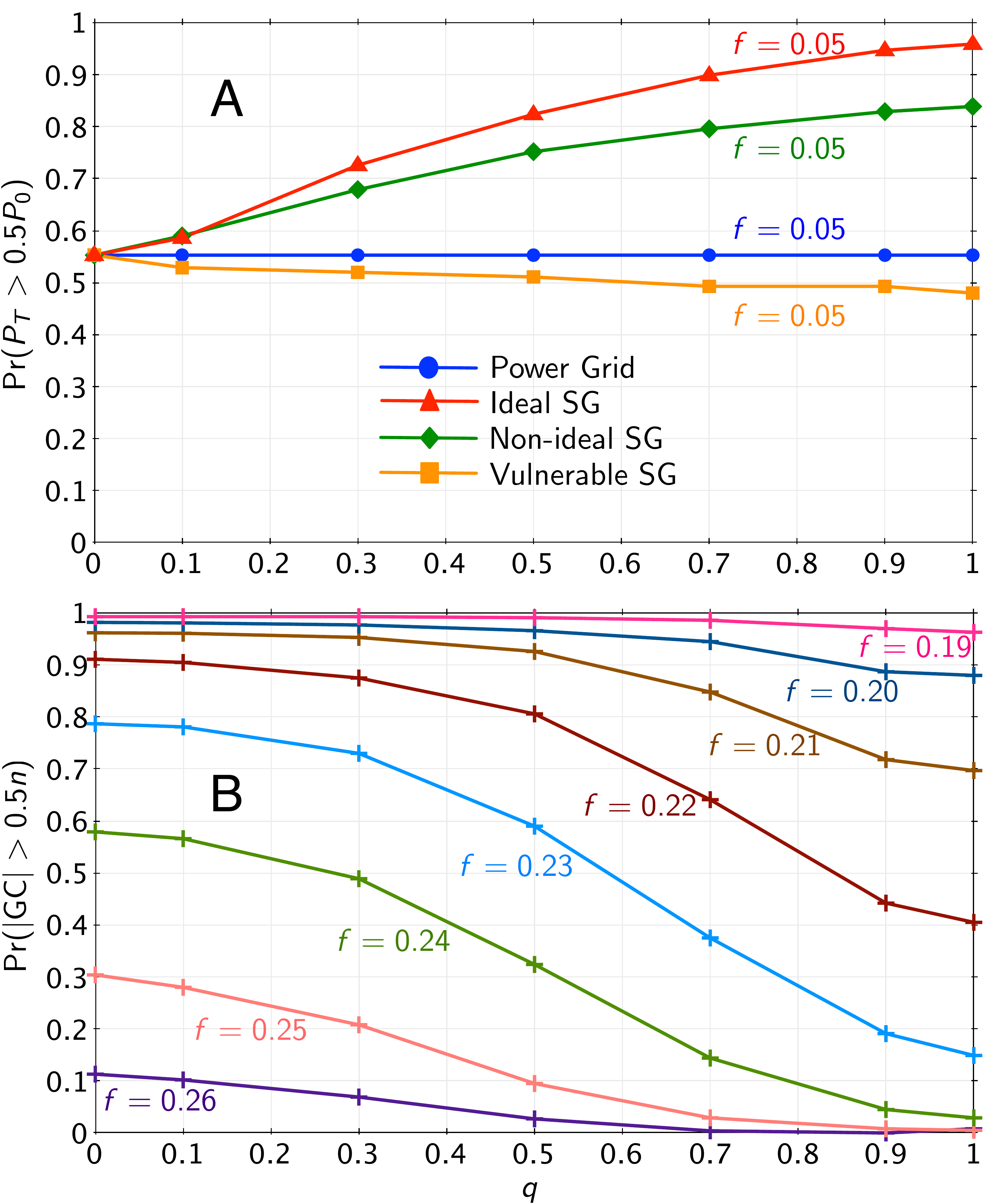}
\caption{Robustness of the Polish network to random failures, with varying levels of coupling,~$q$.
Panel (A) shows results from four different models of cascading in power grids, three of which are coupled to communications systems, after 5\% of nodes initially failed ($f=0.05$).
In this case we measured robustness with the fraction load served after the cascade had subsided ($P_T$).  
Panel (B) reports analogous results from the coupled topological model, for several different failure sizes, with robustness measured as in Figure~\ref{fig:GC_figures_SingleNet}.
}\label{fig:Robustness_vs_coupling}
\end{figure}

Figure~\ref{fig:Robustness_vs_coupling} shows the results for fixed failure sizes,~$f=0.05$, and varying levels of coupling,~$q$.
For~$q=0$ (i.e., uncoupled networks), the smart grid models produce results that are identical to the uncontrolled power grid, since cascading occurs only within the power grid and the communication network neither benefits nor detriments the system.
As~$q$ increases, the robustness of the Ideal and Non-ideal Smart Grid models increase monotonically. 
For the Vulnerable Smart Grid model, robustness decreases monotonically with~$q$.
In contrast, for the Coupled Topological model, robustness decreases monotonically with~$q$; 
the ``optimal'' level of coupling is~$q=0$ for all initiating failure sizes, $f$.
It is interesting to note that the results from both types of model contrast with the results in~\cite{Brummittetal12},
which suggest that there exists an optimal level of coupling between~$q=0$ and~$q=1$.
In all of these cases optimal performance results at either~$q=0$ or~$q=1$.

In order to compare the Non-ideal Smart Grid model to the Coupled Topological model in more detail for different types of topological structures, we took the four additional network topologies from Figure~\ref{fig:GC_figures_SingleNet}, and connected them to correlated communication networks, using the same method used with the Polish power network.
Both models, for~$q=1$, were subsequently subjected to random node failures as before, measuring the robustness of the networks to different disturbance sizes (with varying~$f$).

\begin{figure}[ht]
\centering \includegraphics[width=3.5in]{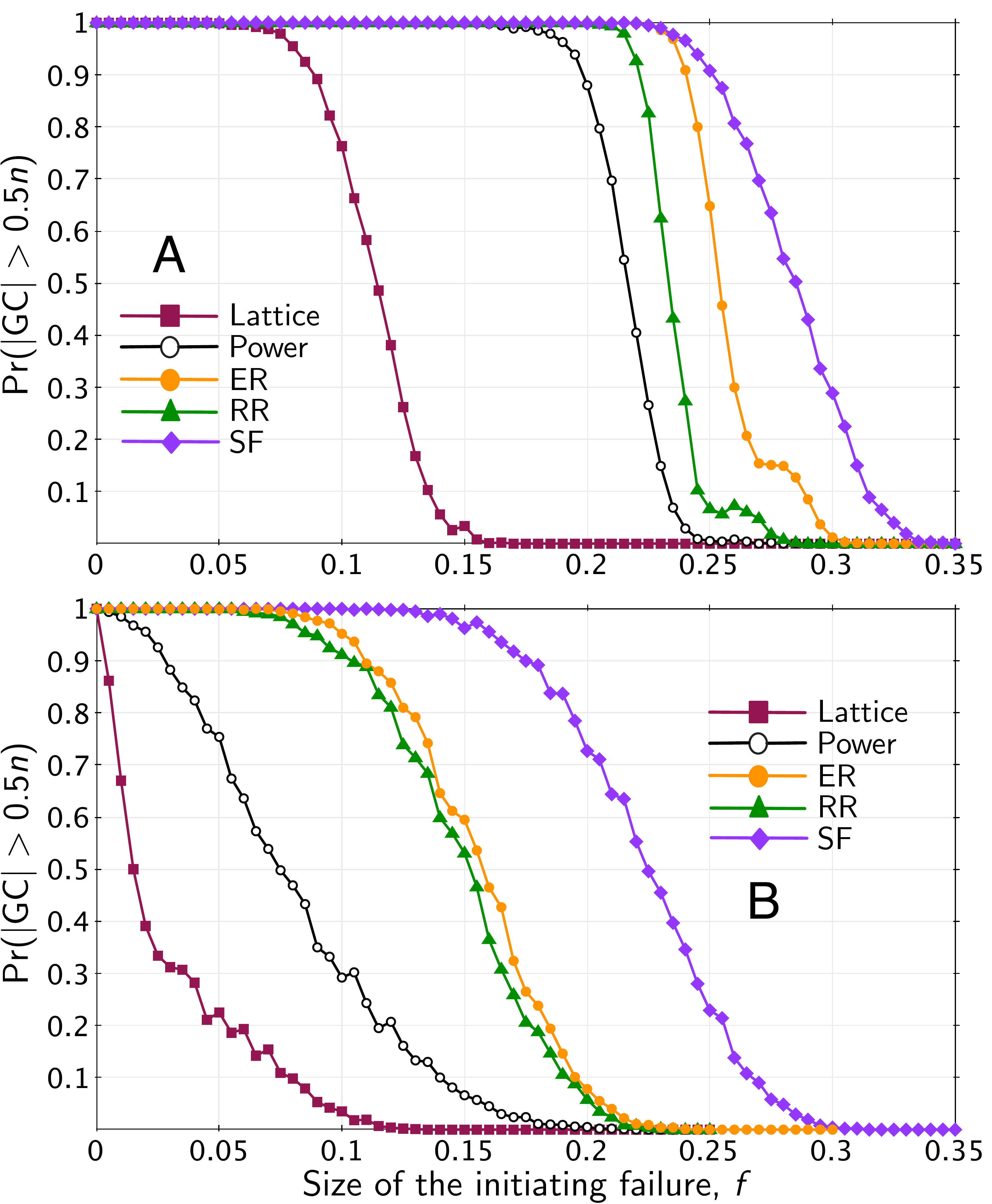}
\caption{Robustness of fully coupled networks, $q=1$, to random failures in the 
(A) Coupled Topological cascading model and in the 
(B) Non-ideal Smart Grid model.}
\label{fig:GC_figures_CoupledNet}
\end{figure}

Figure \ref{fig:GC_figures_CoupledNet} shows the results.
In all five networks, the Coupled Topological model indicates that interdependency increases vulnerability relative to the simple contagion model.
For the Non-ideal Smart Grid model, interdependency decreases vulnerability in every network, relative to the uncoupled Power Grid model in Figure~\ref{fig:GC_figures_SingleNet}.

\section*{Discussion}

Together, these results have important implications both for the emerging science of interdependent networks and for the design of intelligent infrastructure systems.

Firstly, the power grid and topological models show several important qualitative similarities.
The relative vulnerability of the different network structures to random failures is similar across the various models studied in this paper.
Lattices are consistently the most vulnerable and scale-free networks are consistently the most robust;
power grids perform only slightly better than lattice topologies.
This indicates that topological structure does have an important impact on the vulnerability of power networks, and that some aspects of this impact are captured in simple topological models of cascading.

However, this is where the similarities end.  
When we measured the effect of network coupling on performance, increased coupling consistently increased network robustness in all but the most extreme (and unrealistic) power grid model.
For the Ideal and Non-ideal Smart Grid models, the most robust configuration was the fully coupled case,~$q=1$.
In the Coupled Topological model,~$q=0$ was the optimal level of coupling, and robustness monotonically decreased with increased coupling. 
For every attack size, and every topological structure, interdependency increased vulnerability in the coupled topological model and decreased vulnerability in the more realistic smart grid models.
The reason that vulnerability decreased in the smart grid models is that interconnections between the two networks performed valuable functions in arresting the spread of cascades. 
When components were overloaded, and thus at risk of cascading, the communication network facilitated valuable system-wide control functions.
Since these beneficial functions of the communication network are not modeled in the coupled topological model, coupling tends to increase vulnerability. 
These differences indicate that models of network interdependency can lead to misleading conclusions if they do not adequately describe the beneficial functions of coupling in addition to modeling the various mechanisms by which cascades can propagate between the coupled systems.

Finally, the results suggest good design practices for intelligent cyber-physical systems, such as smart grid technology. 
In the case of the Ideal and Non-ideal Smart Grid models, increased coupling was beneficial because of the limited ways in which cascades could propagate between the two networks.
In practice, limits on inter-network cascades can be implemented by sound engineering practices that reduce the chance of failures propagating between networks. 
Adding reliable, well-maintained backup power systems to critical components is a well-known strategy for reducing harmful interdependency.
Most modern SCADA communications systems used in modern power systems have a battery backup power source.
Another example of a useful decoupling strategy is to add battery backup to traffic signals along critical transportation corridors to reduce coupling between power and transportation networks~\cite{Talukdar:2003}.

\section*{Materials and Methods}

\subsection*{Experimental Design}
This section provides detailed descriptions of the assumptions underlying our two experiments.
To summarize:
the objective of Experiment One was to understand how different topological structures respond to random failures, given different models of cascading failure propagation.
The objective of Experiment Two was to compare the network vulnerability, given different models of how cascades propagate between a communications and a power network.
In both cases we were particularly interested to understand similarities and differences between the vulnerability implications of simple topological models and models that more realistically describe the physics of flows in a power grid.

\subsubsection*{Network topological data}
In this study, five different topological structures were studied.
Power network data came from a model of the Polish power grid that is publicly available with MATPOWER~\cite{Zimmermanetal11}.
This model has~$n=2383$ nodes (buses) and~$m=2886$ edges (transmission lines or transformers), after removing parallel edges.
For comparison, four synthetic networks were generated according to the standard
Erd\H{o}s-R\'{e}nyi (ER)~\cite{ErdosRenyi59}, 
random regular (RR)~\cite{Bollobas98}, 
preferential attachment (scale-free, SF)~\cite{BarabasiAlbert99}, and 
square lattice attachment kernels~\cite{Lietal12}.
In order to ensure that the synthetic graphs had the same size as the power network, we randomly removed edges from the initial topological configurations as needed to produce graphs with the correct size.
Edge removals that would result in the graph separating into non-connected subgraphs were avoided in order to ensure that the pre-disturbance graphs were fully connected.
Similarly, duplicate edges and self-loops were removed for consistency with the power grid data.

\subsubsection*{Generating synthetic power grid data}

After building graphs that were identical in size to the 2383-node Polish power grid,
we generated synthetic power grid data for each of the synthetic graphs.
In order locate sources and sinks within the synthetic networks, each of the generators and loads in the Polish network was randomly assigned to one node in each network.
In addition each edge (transmission line)~$i \leftrightarrow j$ was given a normalized impedance of 1, such that the power flowing from~$i$ to~$j$, after our linearized dc power flow assumptions (SM Text), was 
$P_{ij} = \theta_i - \theta_j$,
where~$\theta_x$ is the phase angle of the sinusoidal voltage at node~$x$.
Flow limits on each transmission line were determined by taking the flow limits from with the original Polish network data and randomly assigning each limit to one of the links in the synthetic network. 
After this was done the line limits were increased as needed to ensure that no single line outage would result in a cascading failure, as is common practice in power systems.

\subsubsection*{Generating communications network topologies}

Geographically correlated communication network, $\mathcal{N}_C$, data were generated as follows.
First, we made a copy of the corresponding power network such that~$\mathcal{N}_C = \mathcal{N}_P$.
Then, we randomly rewired 10\% of the endpoints in~$\mathcal{N}_C$, excluding rewirings that would self-loops or duplicate edges.
Then nodes in the two parallel networks were interconnected.
Specifically, node~$i$ in~$\mathcal{N}_P$ was connected to node~$i$ in~$\mathcal{N}_C$ with probability~$q \in [0,1]$. 
The resulting interlinks produce a correlated pair of graphs 
(as illustrated in Figure \ref{fig:coupled_models}), which are at least somewhat similar to the correlated topologies found in real power and communication networks.



\subsection*{Modeling cascading failures in power grids}
Our model of cascading failure in power systems (DCSIMSEP) is based on the model in~\cite{EppsteinHines12}, and is similar to models in~\cite{Dobsonetal07,Pfitzner:2011,Bernstein:2014}, 
which are subsequently closely related to the random fuse networks studied in~\cite{de1985random}.
In this model power flows are computed using the dc power flow equations (SM Text).
The dc model can be summarized as follows:
\begin{align}
    \mathbf{P}_G - \mathbf{P}_D &= \mathbf{B} \boldsymbol{\theta} \label{dcpf1}\\
    f_{ij} &= \frac{1}{x_{ij}} (\theta_i - \theta_j) \label{dcpf2}
\end{align}
where~$\mathbf{P}_G$ and~$\mathbf{P}_D$ are vectors of power generation and load; 
$\mathbf{B}$ is a weighted Laplacian matrix that encodes the network's topology; 
$\boldsymbol{\theta}$ is a vector of voltage phase angles; 
$F_{ij}$ is the power flow from node~$i$ to~$j$; 
and~$x_{ij}$ is the (normalized) inductance of the transmission line.
When a component fails, flows are re-computed according to eqs.~(\ref{dcpf1}) and (\ref{dcpf2}). 
If the revised power flows exceed the flow capacity, this line will open (disconnect) in an amount of time that is proportional to the overload.
This changes the configuration of the network (changing~$\mathbf{B}$), causing the flows to be re-computed.
If the network separates into islands, there may not exist a feasible solution to eq.~(\ref{dcpf1}) due to an imbalance between supply and demand.
To correct this imbalance, a combination of generator adjustments and load reductions are used to arrive at a new, feasible solution of eq.~(\ref{dcpf1}).

\subsection*{Smart Grid Models}

The three smart grid models each depend on an optimization problem that identifies control actions (load shedding and generator reductions) in order to mitigate overloads on transmission lines.
This problem seeks to minimize the amount of load shedding and power generation reductions necessary to arrive at a feasible solution to eqs. (\ref{dcpf1}) and (\ref{dcpf2}), with the added (soft) constraint that each flow~$f_{ij}$ should be within the flow capacity limits for this link.
The optimization proceeds as follows.
In the three smart grid models, we located a ``control center'' at the node in~$\mathcal{N}_C$ with the highest betweenness centrality.
Then, after each 1 minute of simulation time, the control center collects measurement data (power flows as well as generator and load states) from all of the nodes for which there exists a connected path between the control center and that node.
Let~$M$ represent this set of measurable nodes and edges, 
$\overline{M}$ represent the unmeasurable nodes,
$\mathbf{f}_M$ represent the vector of measured power flows,
$\mathbf{P}_{G,M}$ represent the vector of measured generator states, and
$\mathbf{P}_{D,M}$ represent the vector of measured load states.
For the Ideal Smart Grid model,~$\mathbf{f}_M$,~$\mathbf{P}_{G,M}$, and~$\mathbf{P}_{D,M}$ are always full vectors of all measurements from nodes that have communication network connectivity (given~$q$).
Depending on the level of coupling $q$ and the state of the communication network $\mathcal{N}_C$, these may be sub-vectors of all possible measurements.
After the control center gathers measurements~$\mathbf{f}_M$ through the communication system, it solves the following optimization problem:
\begin{align}
\underset{\Delta \mathbf{P}_{D}, \Delta \mathbf{P}_{G}}{\textrm{minimize}} & 
\quad \mathbf{-1^{\boldsymbol{\top}}}\,\Delta \mathbf{P}_{D} +
\boldsymbol{\lambda}^{\boldsymbol{\top}}\,\mathbf{f}_{\textrm{over}} \label{obj}\\
\textrm{subject to}  
& \quad \Delta\mathbf{P}_{G} - \Delta\mathbf{P}_{D} = \mathbf{B}\,\Delta \boldsymbol{\theta}  \label{eq:power_flow}\\
 & \quad  \Delta\theta_{i}=0,\;\forall i\in\Omega_{\textrm{ref}}\label{eq:ref}\\
& \quad  \Delta f_{ij} = \frac{1}{x_{ij}} (\Delta \theta_i - \Delta \theta_j),\,\forall ij \in M \label{eq:flow_eq} \\
& \quad | \mathbf{f}_{M} + \Delta \mathbf{f} | \leq 
    \mathbf{f_{\max}+f}_{\textrm{over}}  \label{eq:flow_limit}\\
 & \quad  \mathbf{f}_{\textrm{over}}\geq \mathbf{0}\label{eq:overload_pos} \\
 & \quad -\mathbf{P}_{G,M} \leq  \Delta \mathbf{P}_{G,M} \leq \mathbf{0} \label{eq:gen_limit}\\
 & \quad -\mathbf{P}_{D,M} \leq  \Delta \mathbf{P}_{D,M} \leq \mathbf{0} \label{eq:demand_limit} \\
 & \quad  \Delta \mathbf{P}_{G,\overline{M}} = \mathbf{0}, 
   \quad \Delta \mathbf{P}_{D,\overline{M}} = \mathbf{0} \label{eq:not_meas_lims}
\end{align}
The objective for this problem (eq.~(\ref{obj})) is to minimize the total amount of load shedding 
($\mathbf{-1^{\boldsymbol{\top}}}\,\Delta \mathbf{P}_{D}$)
plus the weighted sum of all overloads that cannot be eliminated through changes to generators and loads
($\boldsymbol{\lambda}^{\boldsymbol{\top}}\,\mathbf{f}_{\textrm{over}}$).
For this work, we set~$\boldsymbol{\lambda}$ to be uniform weight vector such that each~$\lambda_i = 100$ (in normalized units).
Constraint (\ref{eq:power_flow}) enforces that the net changes to nodal power injections
($\Delta \mathbf{P}_{G} - \Delta \mathbf{P}_{D}$)
must be equal to the changes in power flowing out through transmission lines 
($\mathbf{B}\,\Delta \boldsymbol{\theta}$).
Constraint~(\ref{eq:ref}) fixes one voltage phase angle~$\theta$ in each connected component of the network as a reference;~$\Omega_{\mathrm{ref}}$ represents this set of reference nodes.
Constraint~(\ref{eq:flow_eq}) computes the changes in flow on each of the measured transmission lines.
Eq.~(\ref{eq:flow_limit}) attempts to limit the post-optimization power flows
($\mathbf{f}_{M} + \Delta \mathbf{f}$)
to be below the flow limits,~$\mathbf{f_{\max}}$.
The vector~$\mathbf{f}_{\textrm{over}}$ in eqs.~(\ref{obj}) and~(\ref{eq:flow_limit}) turns the flow constraint into a soft constraint, which alleviates the problem of occasionally infeasible cases, particularly when the system is very heavily stressed.
$\mathbf{f}_{\textrm{over}}$ is constrained to be non-negative in eq.~(\ref{eq:overload_pos}).
Constraints~(\ref{eq:gen_limit}) and~(\ref{eq:demand_limit}) ensure that the system exclusively reduces load and generation at measured nodes ($M$) in its attempt to eliminate overloads on transmission lines.
Finally, eq.~(\ref{eq:not_meas_lims}) forces the system to not change load or generator at nodes that are not accessible from the control center ($\overline{M}$). 
Note that the assumptions used in this model are similar to those presented in~\cite{Parandehgheibietal14}.

Each of the three Smart Grid models makes use of this optimization problem in a slightly different way.
The Ideal Smart Grid model uses perfect information about all communication-connected nodes to solve this problem, optimally choosing adjustments to the available generators and loads, independent of where they are in the network. 
If there is no communication link to a particular node, the Ideal Smart Grid model does not gather data about flows from this location, 
and assumes that it has no ability to control generators or loads at this node.
Thus, the topology of~$\mathcal{N}_C$ does not impact the ideal model.

The Non-ideal Smart Grid model, however, does rely on the state of the communication network.
The optimizer can only control and monitor nodes when there is a $\mathcal{N}_C$ path between a particular grid node and the control center node.
When the path to Node~$i$ is broken, the optimization formulation is adapted to exclude generation and load at Node~$i$ from the set of control variables, and it ignores the flow constraints adjacent to~$i$ (e.g., the flow constraint on Edge~$i \rightarrow j$), unless an adjacent node (e.g.,~$j$) is connected to the control center.
In addition, the non-ideal model assumes that if there is load shedding at grid node~$i$, the adjacent communication node will fail with probability that is equal to the fraction of load shedding.

The Vulnerable Smart Grid Model adds to this the rather extreme assumption that if a communication node fails, the generation and load at that node will also fail.

\section*{Statistical Analysis}

\subsection*{Measuring the initiating failure size}

Note that our measure of attack size~$f$, 
as shown in Figures~\ref{fig:GC_figures_SingleNet}, \ref{fig:Robustness_vs_coupling}, and \ref{fig:GC_figures_CoupledNet}, 
is the complement of the notation used in~\cite{Buldyrevetal10} and in a number of other papers on percolation networks.
In our notation,~$f$ represents the size of the initiating attack (or random failure). 
In~\cite{Buldyrevetal10}, the complement is used, in which ~$p$ represents the fraction of the~$n$ nodes in each network that remain in service immediately after an initial, random set of~$f = \sim(1-p)n$ node failures.
$f$ was used, rather than~$p$, for clarity of presentation, particularly for readers who are less familiar with the percolation literature.

\subsection*{Measuring robustness, sample size}

Note that our measure of robustness~$\Pr(|\text{GC}|>0.5n)$ differs slightly from the traditional~$p_\infty$ measure, which is commonly used in the percolation literature and which averages GC sizes across a set of samples.
Since power networks are small, relative to (for example) thermodynamic systems, the underlying rationale for~$p_\infty$ is less robust.
In our judgement, the~$\Pr(|\text{GC}|>0.5n)$ measure more clearly presented the results.
However, we computed results using both metrics and found that the $p_\infty$ measure led one to the same conclusions as reported in this paper.
See the SM Text for a comparison of the results with~$\Pr(|\text{GC}|>0.5n)$ and~$p_\infty$.

In this paper, each estimate of~$\Pr(|\text{GC}|>0.5n)$ comes from the simulation of 1000 random initiating disturbances of size~$f$ and counting the number of cases that result in a cascade with the end-state largest connected component containing at least~$0.5n$ nodes.
This sample size (1000) was found to provide a reasonable balance between variance in this statistic and computational requirements, which were substantial given the more detailed nature of our models. 
To compute the variance, we used standard bootstrapping methods and found the standard deviation of $\Pr(|\text{GC}|>0.5n)$ to be almost universally less than 0.01.

\section*{Acknowledgments}

The authors gratefully acknowledge financial support from the MITRE Corporation, 
the National Science Foundation Award \#ECCS-1254549, and 
the Defense Threat Reduction Agency Basic Research Grant No. HDTRA1-10-1-0088.
In addition, computational resources were provided by the Vermont Advanced Computing Core (VACC) at the University of Vermont, which is supported by NASA (NNX-08AO96G).

The authors gratefully acknowledge the support of B. Rolfe and J. Kreger, as well as helpful comments and feedback from 
G.~Jacyna, M.~Cohen, C.~Moore, C.~Brummitt, and J.~Bagrow.

The authors are solely responsible for this work.

\pagebreak
\appendix

\renewcommand\thesection{S.\arabic{section}}
\renewcommand\thesubsection{\thesection.\arabic{subsection}}
\renewcommand{\theequation}{S.\arabic{equation}}
\setcounter{equation}{0}
\renewcommand{\thefigure}{S.\arabic{figure}}
\setcounter{figure}{0}

\section*{\LARGE%
Supporting Materials Text (SM Text) for \\
``Reducing Cascading Failure Risk by Increasing Infrastructure Network Interdependency''
}

{Mert~Korkali, Jason~G.~Veneman, Brian~F.~Tivnan, Paul~D.~H.~Hines}


\section{DC Power-Flow Model}

In this paper, we made use of the ``dc power flow'' linearization of the full non-linear power flow equations in our model of cascading failures. 
Here, we briefly describe the derivation of this common, although imperfect, simplification.
For a more detailed discussion of the dc power flow equations and their limitations, see~\cite{Stott:2009,GomezExpositoetal09}.

Consider a node (``bus'' in power systems terminology) $f$ that is connected to node $t$ via a transmission line, which has series resistance $r_{ft}$ and reactance $x_{ft} = \omega l_{ft}$, where $\omega$ is the frequency of the sinusoidal current and $l_{ft}$ is the series inductance of the line.
$r$ and $x$ can be combined to form a complex impedance 
$z_{ft} = r_{ft} + jx_{ft}$, in which (by electrical engineering notational tradition) $j=\sqrt{-1}$.
The inverse of this impedance is known as an ``admittance,'' and is defined as follows: $1/z_{ft} = y_{ft} = g_{ft} + jb_{ft}$, where $g$ and $b$ are known, respectively, as the conductance and susceptance of the line.
The sinusoidal voltages at nodes $f$ and $t$ will each have an amplitude ($V$) and a phase shift ($\theta$, relative to some reference), and can thus be represented with complex numbers 
$\tilde V_f = V_f e^{j\theta_f}$ and $\tilde V_t = V_t e^{j\theta_t}$.
With these definitions, we can define the complex current $\tilde I$ and power $\tilde S$ flowing out from $f$ to $t$ as:
\begin{align}
    \tilde{I}_{ft} &= y_{ft}(\tilde V_f - \tilde V_t) \label{Ift} \\
    \tilde{S}_{ft} &= \tilde V_f I_{ft}^* = \tilde V_f (\tilde V_f^* - \tilde V_t^*) y_{ft}^*  \label{Sft}
\end{align}
where $x^*$ indicates the complex conjugate of $x$.
With some manipulation of eqs.~(\ref{Ift}) and (\ref{Sft}), we can find the active ($P$) and reactive ($Q$) power flowing from $f$ to $t$ as follows:
\begin{align}
    P_{ft} &=  V_f^2 g_{ft} - V_f V_t (g_{ft} \cos \theta_{ft} + b_{ft} \sin \theta_{ft}) \label{eq:P} \\
    Q_{ft} &= -V_f^2 b_{ft} - V_f V_t (g_{ft} \sin \theta_{ft} - b_{ft} \cos \theta_{ft}) \label{eq:Q}
\end{align}
where $\theta_{ft} = \theta_f - \theta_t$ is the phase angle difference between $f$ and $t$.
If we assume 
    that the voltage amplitudes $V_f$ and $V_t$ are at their nominal levels,  
    that we have normalized $y_{ft}$ such that this nominal level is 1.0 (common practice), and
    that the resistance $r_{ft}$ is small (nearly zero) relative to the reactance $x_{ft}$ (a reasonable assumption for bulk power systems), 
then $g_{ft} \cong 0$, and $P_{ft}$ becomes:
\begin{align}
    P_{ft} \cong - b_{ft} \sin \theta_{ft} = \frac{1}{x_{ft}} \sin \theta_{ft} \label{eq:P_almost}
\end{align}
If we assume that $\theta_{ft}$ is small, then $\sin \theta_{ft} \cong \theta_{ft}$ and we get:
\begin{align}
    P_{ft} \cong \frac{1}{x_{ft}} \theta_{ft} \label{eq:P_dc}
\end{align}
If we furthermore assume that $Q_{ft}=0$ (not a particularly good assumption), 
then the current magnitude and the power are equal, $|I_{ft}| = P_{ft}$, and we can use eq.~(\ref{eq:P_dc}) to roughly simulate power flows in a power system.

In order to solve for the flows $P_{ft}$ in simulation, we put eq.~(\ref{eq:P_dc}) into matrix form as follows.
Let $\mathbf{A}$ denote the line-to-node incidence matrix with 1 and -1 in each row indicating the endpoints of each line,
$\boldsymbol{\theta}$ be the vector of voltage phase angles,
$\mathbf{X}$ be a diagonal matrix of line reactances, and 
$\mathbf{P_\text{flow}}$ be a vector of active power flows along transmission lines. 
Then, we can solve for the vector of power flows $\mathbf{P_\text{flow}}$ given that we know the vector of voltage phase angles $\boldsymbol{\theta}$ as shown in the following:
\begin{align}
  \mathbf{A}^{\boldsymbol{\top}} \boldsymbol{\theta} &=  \mathbf{X} \mathbf{P_\text{flow}}  \\
   \mathbf{P_\text{flow}} &=  \left[\mathbf{X}^{-1} \mathbf{A}^{\boldsymbol{\top}}\right] \boldsymbol{\theta}
\end{align}
In order to solve for $\boldsymbol{\theta}$, we use information about the sources (generators) and sinks (loads) to build a vector of net injected powers (generation minus load), $\mathbf{P}$.
Given $\mathbf{P}$, we can solve the following to find $\boldsymbol{\theta}$:
\begin{equation}
\mathbf{P} = \mathbf{A} \mathbf{P_\text{flow}} = \left[\mathbf{A} \mathbf{X}^{-1} \mathbf{A}^{\boldsymbol{\top}}\right] \boldsymbol{\theta} =  \mathbf{B} \boldsymbol{\theta}
\end{equation}
The matrix $\mathbf{B}$ is known as the bus susceptance matrix, and has the properties of a weighted graph Laplacian matrix describing the network of transmission lines, where the link weights are the susceptances $b_{ft} = 1/{x_{ft}}$. 

\section{Supplemental results}

\subsection{Comparing $\Pr(|\text{GC}|>0.5n)$ to $p_\infty$}
In this paper, we measured the impact of disturbances of various sizes, $f$, on the probability of at least half of the network remaining within the ``giant component'' after the resulting cascade had subsided $\Pr(|\text{GC}|>0.5n)$, 
or the probability of half of the load still being served after the cascade completed: $\Pr(P_T>0.5 P_0)$.
An alternative way to measure the impact of the disturbances is to measure the average cascade size (sometimes known as the yield), rather than the probability of a cascade in a given size range.
This measure would be more analogous to the $p_\infty$ metric that is commonly used in the literature on phase transitions in percolation systems.
We chose not to use $p_\infty$ as our primary measure of network robustness since the modeling assumptions described in the above discussion of ``dc power flow'' become particularly inaccurate for very large cascades.
Essentially, $p_\infty$ would, in many cases, average over small numbers that were not particular accurate.

However, the results that one obtains by measuring the average cascade impact do not lead one to substantially different conclusions than those reported in the paper (aside from the fact that the transitions are much more gradual).

\begin{figure}[ht]
\centering
\includegraphics[width=3in]{Figures/GC-Figures-5Networks-SingleNetModels-inverted.pdf}
\includegraphics[width=3in]{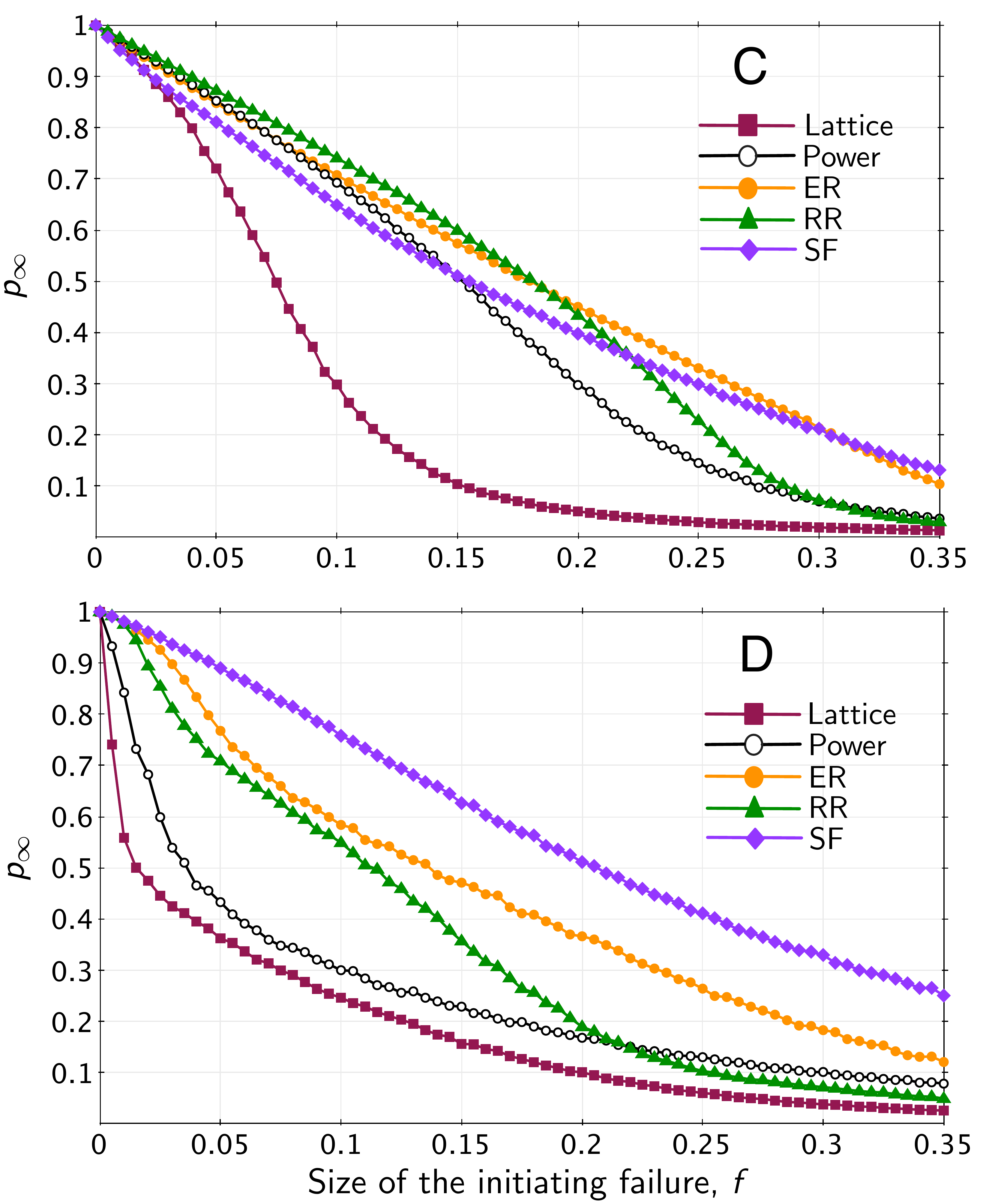}
\caption{The response of a simple model of topological contagion (A, C) and our power grid model (B, D) to random node failures.
Panels A and B show the response using the $\Pr(|\text{GC}|>0.5n)$ measure used in the paper. 
Panels C and D show the response using the average size of the post-cascade giant component, $p_\infty$. 
}
\label{fig:contagion_vs_power}
\end{figure}

Figure~\ref{fig:contagion_vs_power} compares the response of various networks to random failures using the $p_\infty$ and $\Pr(|\text{GC}|>0.5n)$ measures for the topological contagion and power grid models.
For the power grid model, the relative robustness of the five network structures is unchanged. 
The lattice is the most vulnerable and the scale-free network is the most robust.
In the topological model, the $p_\infty$ measure indicates that the power grid, random graph, random regular, and scale-free networks have similar levels of robustness, for $f<0.15$. The lattice remains to be the most vulnerable of the five network structures.

\begin{figure}[ht]
\centering
\includegraphics[width=3in]{Figures/Physics-Topological-vs-coupling_1000avg_p_05.pdf}
\includegraphics[width=3in]{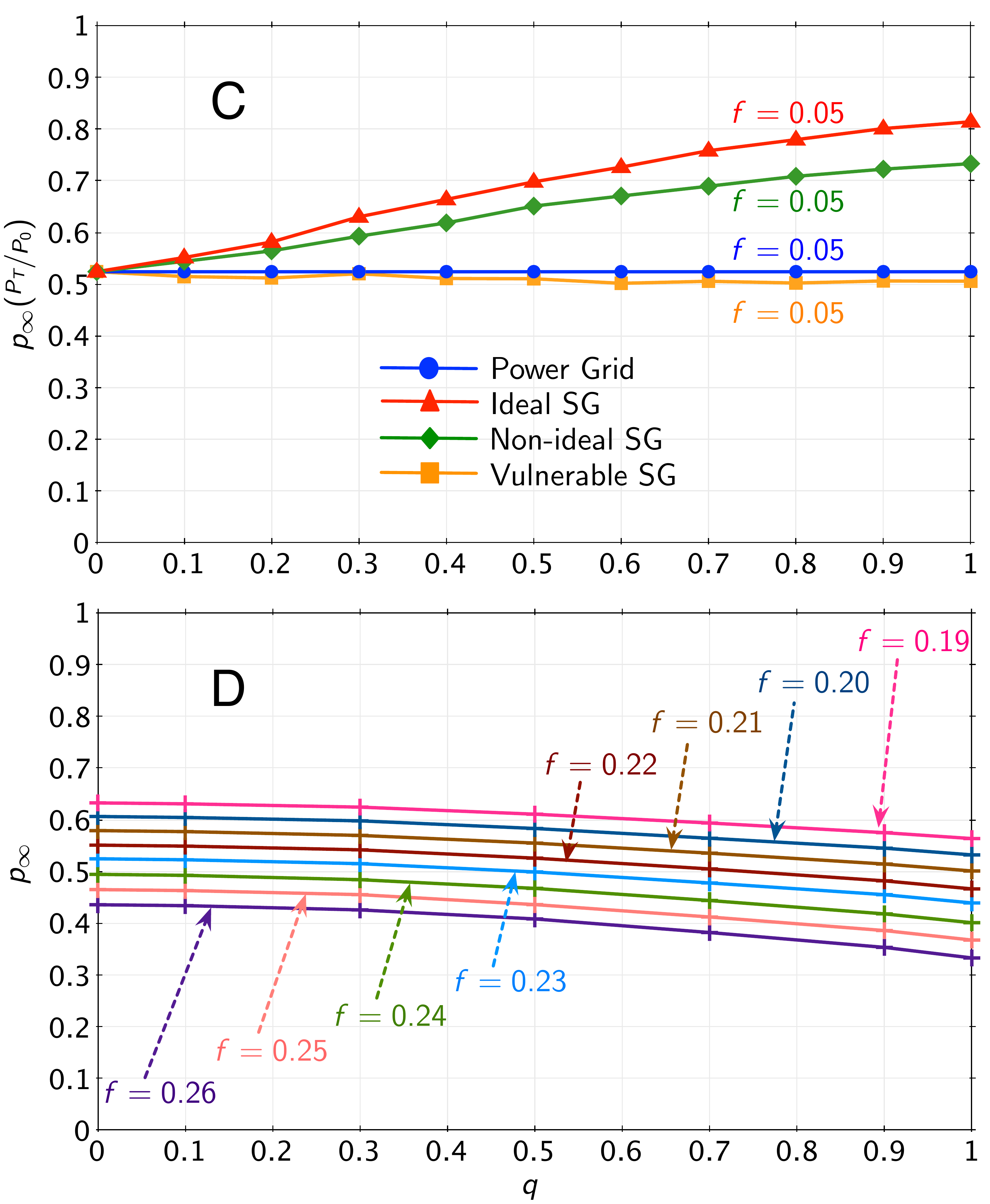}
\caption{The robustness of the Polish power grid topology, when coupled to a communication network, for two different models and two different measures of robustness, as a function of the level of inter-network coupling, $q$.
Panels A and C show results for the four power grid models, whereas B and D show the Coupled Topological model.
Panels A and B measure robustness using the $\Pr(|\text{GC}|>0.5n)$ and $\Pr(P_T>0.5 P_0)$ measures, as in the main paper, whereas C and D use measures that are more analogous to $p_\infty$. }
\label{fig:robustness_and_coupling}
\end{figure}

Figure~\ref{fig:robustness_and_coupling} compares the response of various coupled models to random failures with different levels of coupling between the power and communications network.
In this case, we compare the original metrics used in the paper ($\Pr(|\text{GC}|>0.5n)$ and $\Pr(P_T>0.5 P_0)$) to $p_\infty$.
Our analogous measure of robustness for the four Power/Smart Grid models is $P_T / P_0$: the ratio of the amount of load connected at the end of the cascade to the original load.
The results for the four different (smart) power grid models are not substantially changed.
We still see that increased coupling increases robustness in both the Ideal and the Non-Ideal Smart Grid models, whereas coupling is detrimental (though only slightly) in the Vulnerable SG model.
For the Coupled Topological model, coupling is detrimental to robustness; indeed, by measuring the results using both $p_\infty$ and $\Pr(|\text{GC}|>0.5n)$, the decrease in performance with $q$ is monotonic.

\subsection{Comparisons among the various models}
\begin{figure}[htb]
\begin{center}
\centerline{\includegraphics[width=0.9\columnwidth]{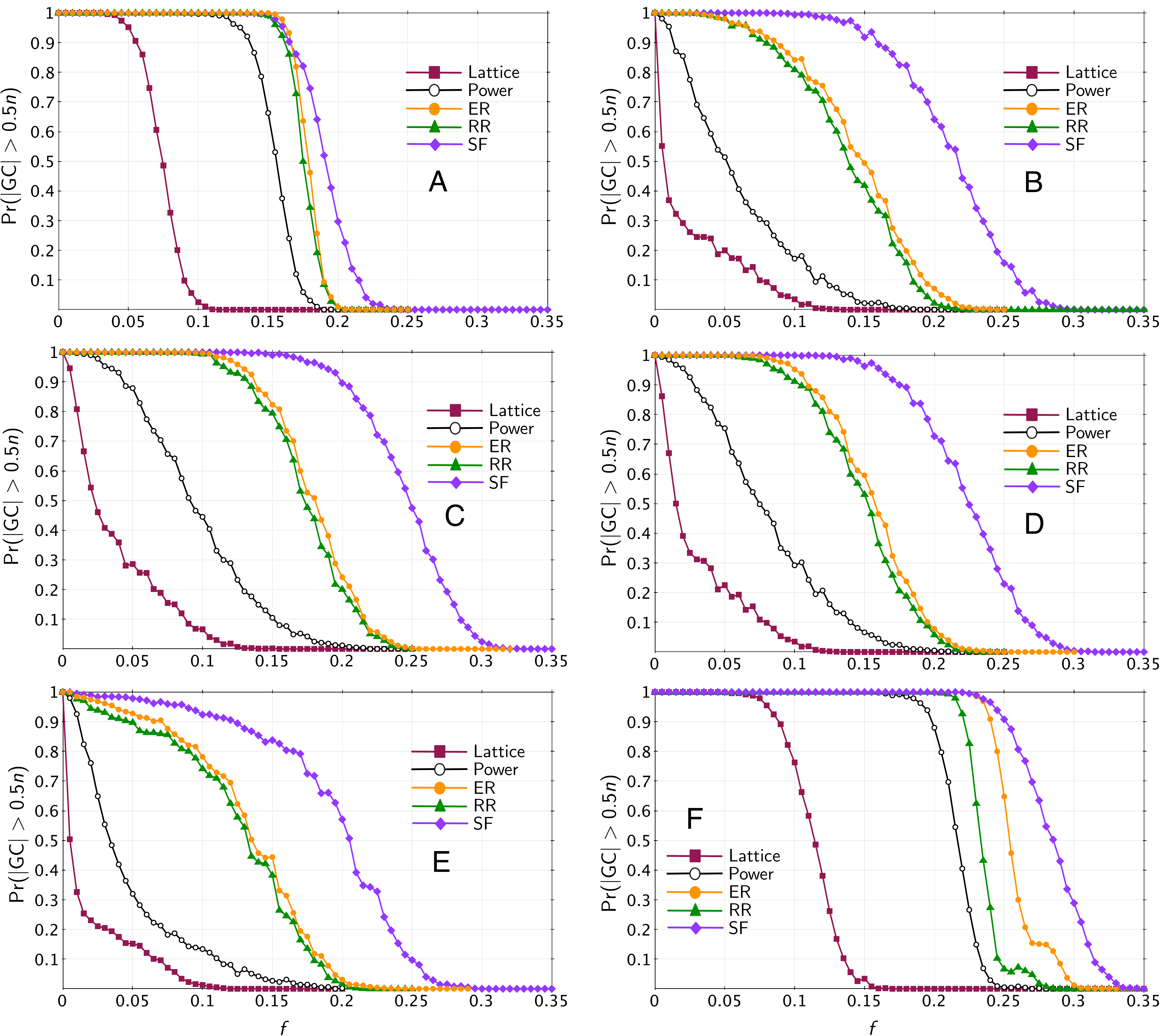}}
\caption{Comparison of the robustness for all six of the models considered in this paper:
(A) single-network, topological cascading, (B) single-network, power-flow-based cascading, (C) ideal smart grid, (D) non-ideal smart grid, (E) vulnerable smart grid, and (F) coupled topological cascading.
In all of the coupled models the grid and communication nodes are assumed to be \textit{perfectly} coupled, i.e., $q=1$.}
\label{fig:GC_figures}
\end{center}
\end{figure}

For ease of comparison among the six different models considered in this paper, Figure~\ref{fig:GC_figures} compares the robustness, measured using $\Pr(|\text{GC}|>0.5n)$ for all six models.
For the coupled network models, these results show the fully coupled case, $q=1$.

\FloatBarrier

\subsection{50\% coupling results}

To better understand the impact of the level of coupling, we re-computed the results shown in Figure~\ref{fig:GC_figures} using 50\% coupling $q=0.5$.

\begin{figure}[hbtp]
\begin{center}
\centerline{\includegraphics[width=0.9\columnwidth]{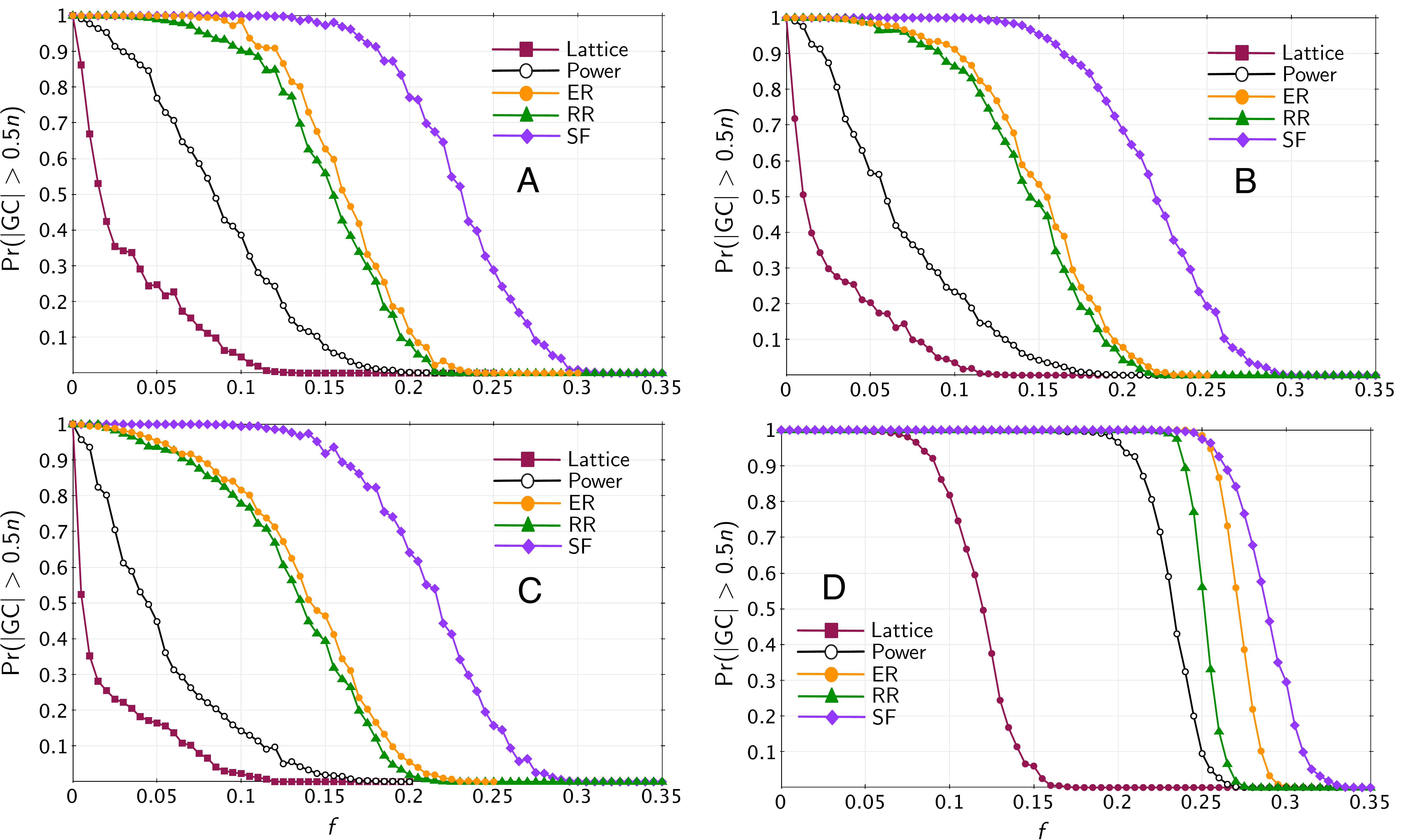}}
\caption{Comparison of robustness results for the four coupled network models, with 50\% coupling, $q=0.5$, for the:
(A) ideal smart grid model, (B) non-ideal smart grid model, (C) vulnerable smart grid model, and (D) Coupled topological cascading model.}\label{fig:GC_figures_HalfCoupling}
\end{center}
\end{figure}

\FloatBarrier

\subsection{Network Vulnerability Indices}

One way to compare the various topological configurations and models described in this paper is to convert the sigmoidal results shown in Figures \ref{fig:GC_figures} and \ref{fig:GC_figures_HalfCoupling} into a single metric of robustness (or conversely, vulnerability).
To quantify the effects of topology, physics, and coupling among different synthetic networks, we define the following network vulnerability index ($\beta$) as follows:
\begin{align}\label{eq:NVI_1}
\beta &= -\log\int_{0}^{1} \mathsf{Pr_{\text{GC}}} (f) \ \mathrm{d}f \\ \label{eq:NVI_2}
  &\approx -\log \bigg\{\frac{1}{2L}  \sum_{\ell=1}^{L-1} \mathsf{Pr_{\text{GC}}}(f_\ell) + \mathsf{Pr_{\text{GC}}}(f_{\ell+1}) \bigg\}
\end{align}
where $f$ is the initiating failure size; 
$L$ is the total number of $f$ values simulated; 
and $\mathsf{Pr_{\text{GC}}}=\Pr(|\text{GC}| > 0.5n)$ is the probability of observing a GC whose size is more than half the number of grid nodes.

\begin{figure}[hbtp]
\centering
\includegraphics[width=1\columnwidth]{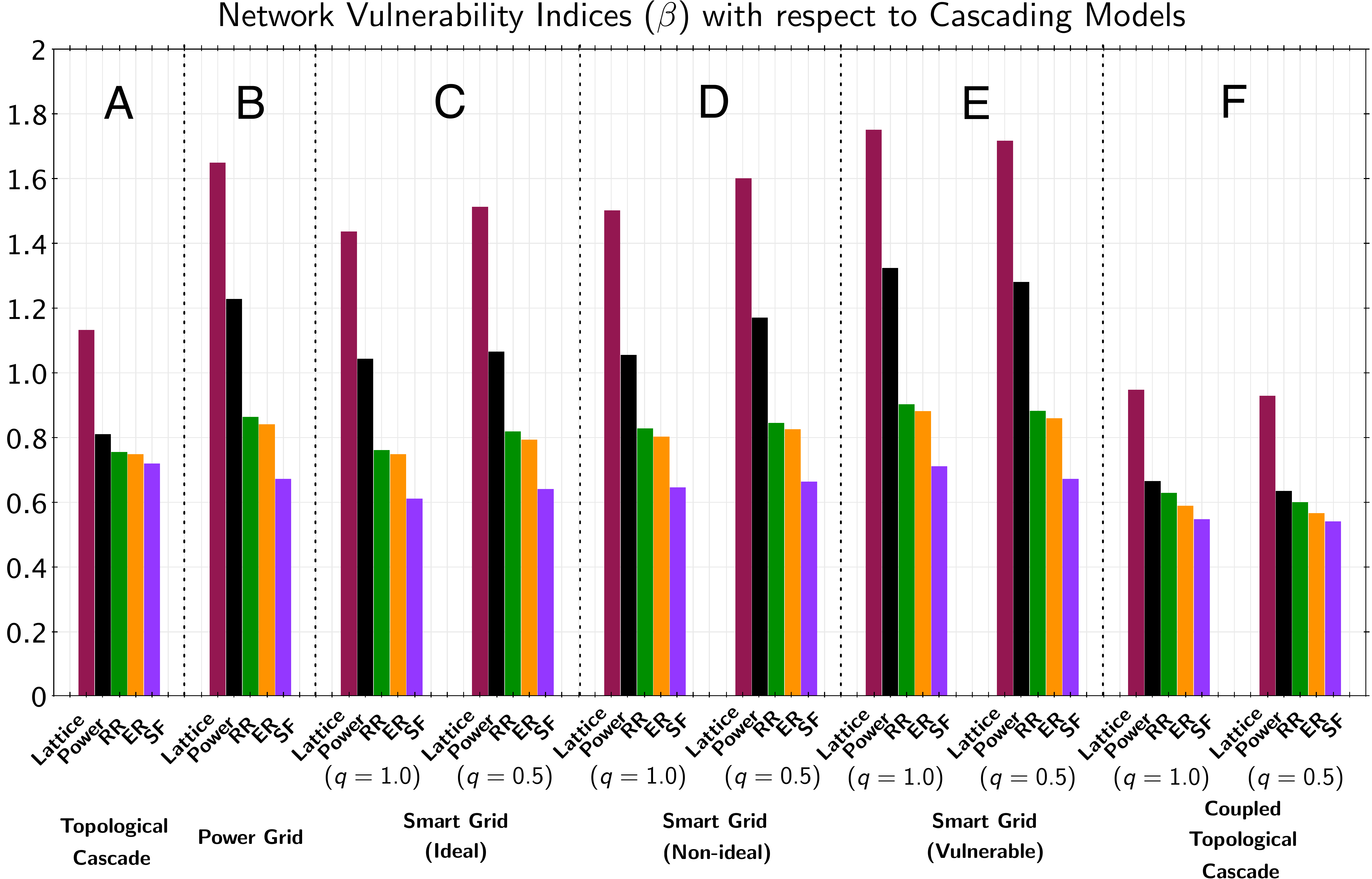}
\vspace{-.4cm}
\caption{Network vulnerability indices, $\beta$ for different models of cascading for the fully coupled ($q=1.0$), and half coupled ($q=0.5$) cases.
Panel A shows the simple topological cascading model;
B shows the uncoupled power grid;
C, D and E show the three smart grid models; 
and F shows the Coupled Topological model.}
\label{fig:vuln_idx_bar_graphs_full_coupling}
\end{figure}

\pagebreak 
\bibliographystyle{ScienceAdvances}
\bibliography{References}


\end{document}